%% file: main.tex
\documentclass[journal]{vgtc}                     

\onlineid{5086}

\ieeedoi{xx.xxxx/xxxx.201x.xxxxxxx}

\vgtccategory{Research}

\vgtcpapertype{please specify}

\title{From Static to Interactive: Authoring Interactive Visualizations\\ via Natural Language}

\author{%
  \authororcid{Can Liu}{0000-0002-1175-0734},
  Jaeuk Lee, Tianhe Chen, Zhibang Jiang,
  \authororcid{Xiaolin Wen}{0000-0002-8562-7640}
  and 
  \authororcid{Yong Wang}{0000-0002-0092-0793}
}

\authorfooter{
  \item
  	C. Liu, T. Chen, X. Wen, and Y. Wang are with Nanyang Technological University.
  	E-mail: can.liu@ntu.edu.sg, \{chen1594, xiaolin004\}@e.ntu.edu.sg, yong-wang@ntu.edu.sg. Y. Wang is the corresponding author.
    
  \item
  	J. Lee is with Ajou University.
  	E-mail: woukl22@ajou.ac.kr.
  
  \item Z. Jiang is with Providence Health \& Services.
  	E-mail: zhibang.jiang@newschool.edu.
}

\usepackage{enumitem}

\setlist[itemize]{noitemsep, topsep=0pt, leftmargin=*}

\newcommand{\revision}[1]{\leavevmode{\textcolor[RGB]{220, 20, 60}{#1}}}
\newcommand{\remark}[1]{\textcolor[RGB]{150, 200, 0}{#1}}

\newcommand{\removed}[1]{\leavevmode{\color{red}{\st{#1}}}}

\def \cleanversion{} 
\ifx\cleanversion\undefined
\else
 \renewcommand{\remark}[1]{\iffalse #1 \fi} 
 \renewcommand{\removed}[1]{\iffalse #1 \fi} 
 \renewcommand{\revision}[1]{#1}
\fi

\newcommand{\toolname}[1]{\textit{Athanor}}

\abstract{Interactivity is crucial for effective data visualizations. However, it is often challenging to implement interactions for existing static visualizations, since the underlying code and data for existing static visualizations are often not available, and it also takes significant time and effort to enable interactions for them even if the original code and data are available. 
To fill this gap,
we propose \toolname{}, a novel approach to transform existing static visualizations into interactive ones using multimodal large language models (MLLMs) and natural language instructions. Our approach introduces three key innovations: (1) an action-modification interaction design space that maps visualization interactions into user actions and corresponding adjustments, (2) a multi-agent requirement analyzer that translates natural language instructions into an actionable operational space, and (3) a visualization abstraction transformer that converts static visualizations into flexible and interactive representations regardless of their underlying implementation.
\toolname{} allows users to effortlessly author interactions through natural language instructions, eliminating the need for programming.
We conducted two case studies and in-depth interviews with target users to evaluate our approach. The results demonstrate the effectiveness and usability of our approach in allowing users to conveniently enable flexible interactions for static visualizations. %
}


\teaser{
  \centering
  \includegraphics[width=\textwidth]{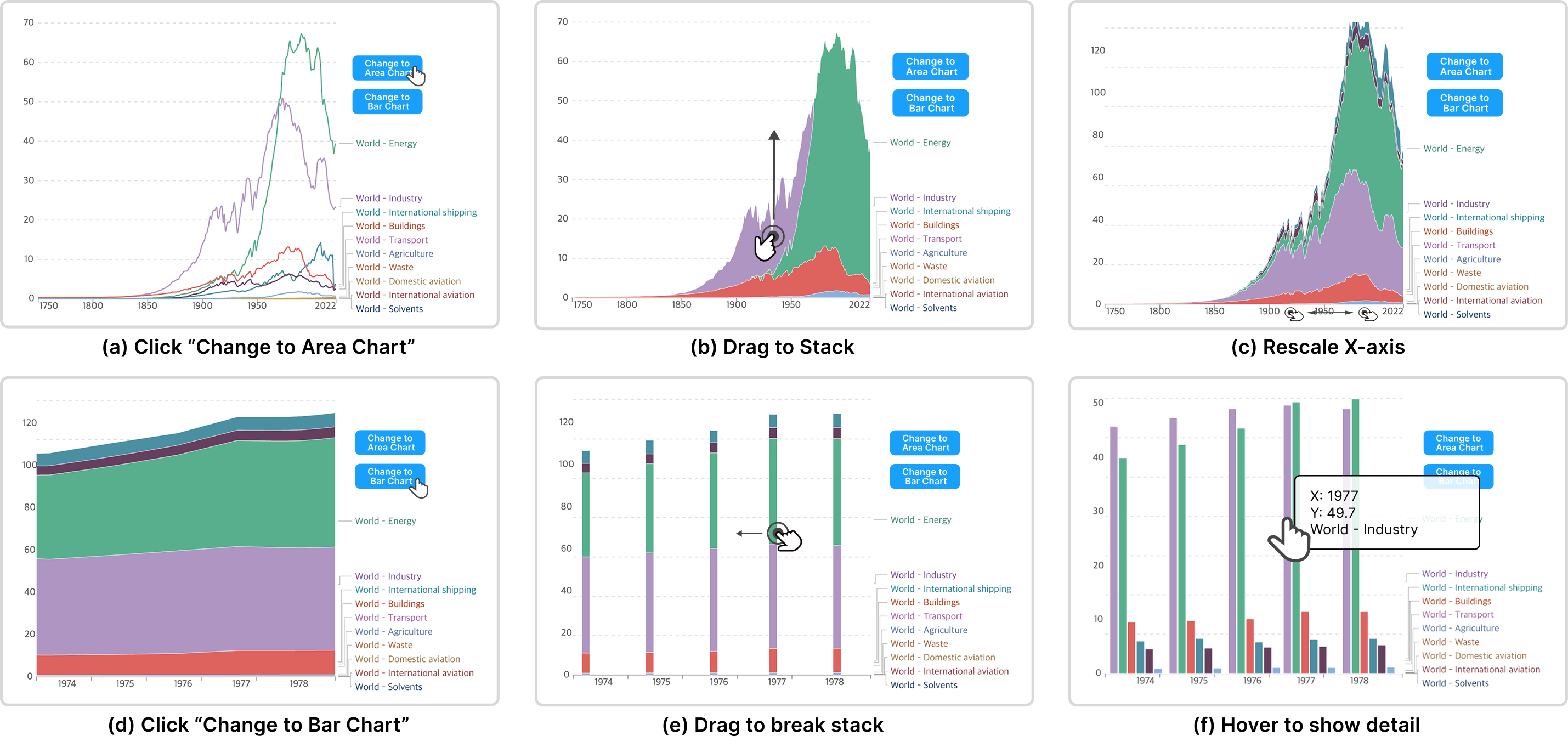}
  \caption{Given a static chart (a), \toolname{} supports enabling interactions for it. The static chart has been enhanced with buttons to modify chart types (transforming (a) to (b) and (d) to (e)), the ability to rearrange the visual elements to stack or group ((b) to (c) and (e) to (f)), rescaling on the x-axis ((c) to (d)), and hover functionality to display a detailed tooltip (f). After authoring, the static line chart can support interactive features that enable users to perform in-depth analysis.}
  \label{fig:bar_chart}
}

\graphicspath{{figs/}{figures/}{pictures/}{images/}{./}} 

\usepackage{tabu}                      
\usepackage{booktabs}                  
\usepackage{lipsum}                    
\usepackage{mwe}                       

\usepackage{mathptmx}                  

\usepackage{CJK}            
\usepackage{listings}

\usepackage{enumitem}
\setlist[itemize]{leftmargin=8pt, itemsep=0pt, topsep=2pt}

\usepackage{_macros}

\renewcommand{\wsicon}[1]{\raisebox{-0.3\height}{\includegraphics[height=1.1em]{image/icon/#1.png}}}

\usepackage{amsmath}
\usepackage{soul}
\usepackage{xcolor}
\setstcolor{red}
\begin{document}
\begin{CJK}{UTF8}{gbsn}  



\maketitle

\input{chapter/1_intro.tex}

\input{chapter/2_related}

\input{chapter/3_overview}

\input{chapter/4_actionspace}

\input{chapter/5_nlinput.tex}
\input{chapter/6_representchart}
\input{chapter/7_evaluation}
\input{chapter/8_discussion}
\acknowledgments{%
This project is supported by the Ministry of Education, Singapore, under its Academic Research Fund Tier 2 (Proposal ID: T2EP202220049), and by the Start Up Grant awarded to Yong Wang. Any opinions, findings and conclusions, or recommendations expressed in this material are those of the author(s) and do not reflect the views of the Ministry of Education, Singapore.
}

\bibliographystyle{abbrv-doi-hyperref}

\bibliography{main}

\clearpage
\begingroup
\setlength{\textfloatsep}{0pt}
\setlength{\floatsep}{0pt}
\setlength{\intextsep}{0pt}

\endgroup

\end{CJK}
\end{document}

%% file: chapter/1_intro.tex
\section{Introduction}

Interactivity adds a new dimension to visualizations, enabling users to dynamically explore data, uncover hidden patterns, and tailor views to their specific needs.
For instance, interactive visualization software like Tableau and Power BI enable real-time filtering and zooming, revealing insights that are typically inaccessible in static visualizations. 
However, despite the power of these interactive visualization tools, static visualizations remain prevalent in published outputs~\cite{weissgerber2016static}, typically presenting data from a fixed perspective with underlying data and code usually inaccessible to viewers. 
This static nature limits users' ability to adjust focus, compare different time periods, and segment data, thereby hindering users from gaining deeper data insights.
Interactivity can significantly enhance the utility of static visualizations, yet integrating it into static formats is far from being straightforward and easy~\cite{snyder2023divi}. 
There are three major challenges that have limited the adoption of interactive visualizations:



\begin{itemize}[noitemsep,topsep=2pt,parsep=0pt,partopsep=0pt]
\item \textbf{C1. Interaction Complexity:}
The interaction space is vast as it encompasses multidimensional design space ranging from different actions (e.g., clicking, hovering, dragging) to various analytical intents (e.g., filtering, comparing, and hypothesis validation)~\cite{yi2007toward, heer2012interactive}. 
It is often time-consuming to enable interactions for a static visualization.
Even with access to the original code and data, modifying each visualization's code to add interactive features requires substantial visualization expertise and implementation skills.

\item \textbf{C2. User Requirement Diversity:}
The space of user-required interaction functionalities is enormous.
Different users have distinct requirements based on their specific scenarios, tasks, and analysis goals.
For example, data analysts may need complex filtering and comparison interactions, while casual users prefer simple highlighting and annotation features.
This diversity in user requirements across different contexts makes it challenging to provide a standardized implementation of visualization interactions.

\item \textbf{C3. Visualization Implementation Diversity:}
There are various types of visualizations (e.g., bar charts, area charts, line charts, etc.) and visualizations 
can be implemented using different toolkits, e.g., Vega-Lite~\cite{satyanarayan2017vegalite}, ECharts~\cite{li2018echarts}, and D3~\cite{bostock2011d3}, resulting in substantially different code implementations.
This heterogeneity makes it difficult to enable interactions for visualizations implemented by different visualization toolkits.
Furthermore, many existing visualizations lack access to the underlying data and code, making it infeasible to support interactivity by extending the original implementation.
\end{itemize}

To address these challenges, we propose \toolname{}, a novel approach that allows users to \textbf{add interactions to} \textbf{existing static visualizations conveniently and efficiently} based on users' natural language inputs.
Our approach comprises \textit{an action-modification interaction design space},
\textit{a multi-agent requirement analyzer}, and \textit{an implementation-agnostic representation translator}.
The \textbf{action-modification interaction design space} maps interactions to \textit{user actions} and their corresponding \textit{visualization modifications}, providing a structured framework to describe the potential interactions supported by a static visualization's information. This design space specifies how user actions trigger specific visualization modifications, forming an actionable interaction model (\textbf{C1}).
The \textbf{multi-agent requirement analyzer} interprets users' natural language requirements, transforming them into actionable specifications within the action-modification design space.
Through a collaborative agent-based approach, it processes diverse inputs to precise operations by translating inputs into structured sequences of actions and modifications, correcting parsing errors, and providing guidance to refine unclear requests (\textbf{C2}).
The \textbf{visualization abstraction translator} converts existing visualizations into a flexible abstraction independent of the underlying implementation (\textbf{C3}).
The translator leverages multimodal large language models (MLLMs) to parse existing visualizations, representing the results through a constraint model~\cite{liu2024spatial}, which is an implementation-agnostic representation for existing visualizations.



\toolname{} can effectively parse the content of existing visualizations in SVG format, especially the most popular charts (e.g.,  bar charts, line charts, scatter plots, and area charts), and then directly enable interactivity on them, which can maintain the visual consistency between the interactive visualizations and the original static visualizations.
To demonstrate the effectiveness of \toolname{}, we first 
present two case studies to illustrate how users can conveniently enable interactions for a given visualization with the help of our approach.
Also, we conducted in-depth user interviews with 11 participants. The feedback indicates that our action-modification design space can cover various user requirements, and \toolname{} allows users to effectively add interactions to static visualizations with natural language inputs.
The contributions of this work are as follows:
\begin{itemize}[noitemsep,topsep=2pt,parsep=0pt,partopsep=0pt]
\item We propose \toolname{} that consists of three major components: action-modification interaction design space, multi-agent requirement analyzer, and visualization abstraction translator.
It allows users to conveniently author flexible interactions for static visualizations via natural language interaction.

\item We present two case studies and conduct in-depth user interviews to demonstrate the effectiveness and usefulness of \toolname{}.
A gallery of interaction-enhanced visualizations created by using our approach can be found at 
\textcolor{blue}{\url{https://authorvisinter.github.io}}.

\end{itemize}

%% file: chapter/2_related.tex

\section{Related Work}
\label{section:background}

Our work provides a natural language interface (NLI) for authoring interactions on static visualizations, which relates to visualization interaction, NLI for visualization, and visualization enhancement.

\subsection{Visualization Interaction}

Visualization interaction encompasses various ways users engage with and manipulate visual representations of data.
We treat a visualization interaction as user actions and visualization modifications.

\textbf{Action-based interactions.}
Several researchers have categorized visualization interactions based on user actions performed via input devices such as mice, keyboards, and touchpads.
Shneiderman~\cite{shneiderman2003eyes} outlines WIMP (windows, icons, menus, pointer) actions to present users' interaction intentions, such as clicking to zoom and typing to filter, emphasizing direct manipulation with mouse and keyboard inputs.
Expanding on this, Ward et al.\cite{ward2010interactive} proposed a taxonomy of actions, including pointing (e.g., mouse clicks to select), dragging (e.g., mouse movements to resize or reposition), scrolling (e.g., mouse wheel or touchpad gestures to navigate), and keypressing (e.g., keyboard shortcuts to toggle views).
These physical action frameworks have influenced interactive visualization systems. For example, Polaris\cite{stolte2002polaris} maps mouse-based actions (e.g., clicking to highlight, dragging to brush, or context-clicking to drill down) to specific responses.
Several approaches~\cite{Baur2012touchwave, ramik2018touchdata} studied touchpad interactions, identifying multi-touch gestures like pinching to zoom or swiping to pan as actions in visualization systems.

\textbf{Modification-based interactions.}
Another perspective on visualization interaction focuses on how visualization modifies its presentation to achieve users' goals.
Yi et al.~\cite{yi2007toward} categorized visualization interactions into several types: select (marking something as interesting), reconfigure (showing a different arrangement), encode (giving a different representation), abstract/elaborate (showing more or less detail), filter (showing something conditionally), and connect (showing related items).
Brehmer and Munzner~\cite{brehmer2013multi, munzner2014visualization} classified visualization modifications into six categories: selecting, navigating, arranging, changing, filtering, and aggregating.
Sedig and Parsons~\cite{sedig2013interaction} offered an alternative taxonomy distinguishing between unipolar and bipolar interactions.


In our work, we define an interaction as comprising both user actions and the corresponding visualization modifications.
We categorize user actions according to different input devices (e.g., mouse, touchpad, or keyboard) and classify modifications based on the possible operations that can be performed on an existing visualization.

\subsection{NL Interface for Visualization}

Over the past decades, the field of natural language interfaces (NLIs) for visualization~\cite{luo2021nl2vis, luo2018deepeye, shen2022towards, wang2022towards, wang2023data, liu2021advisor} has attracted significant attention.

\textbf{NLI for constructing visualizations.}
Cox et al.~\cite{cox2001multi} introduced the first pipeline for constructing visualizations from natural language input.
To address the inherent ambiguity of natural language, machine learning (ML)-based methods~\cite{sun2010articulate} and interactive-based methods~\cite{gao2015datatone} have been proposed.
Articulate~\cite{sun2010articulate} employs machine learning techniques to classify visualization tasks based on the words in sentences. DataTone~\cite{gao2015datatone} introduces ambiguity widgets that facilitate interaction to reduce ambiguity.
Shi et al.~\cite{shi2021talk2data} proposed a method for decomposing high-level questions into simpler components, enabling the generation of visual answers.
Previous methods primarily rely on rule-based or learning-based techniques at the word level, but they are limited in handling the large diversity present at the sentence level.
In recent years, large language models (LLMs)~\cite{wolf2020transformers} (e.g., BERT~\cite{devlin2018bert}, T5~\cite{raffel2020exploring}, ChatGPT~\cite{openai_chatgpt_api}) have demonstrated impressive capabilities for adapting to visualization generation tasks~\cite{li2024visualization, kavaz2023chatbot}.
For example, Maddigan et al.\cite{maddigan2023chat2vis} and Luo et al.~\cite{luo2021natural} translate natural language to visualization specificatio
\begin{figure*}[!t]
    \centering
    \includegraphics[width=\textwidth]{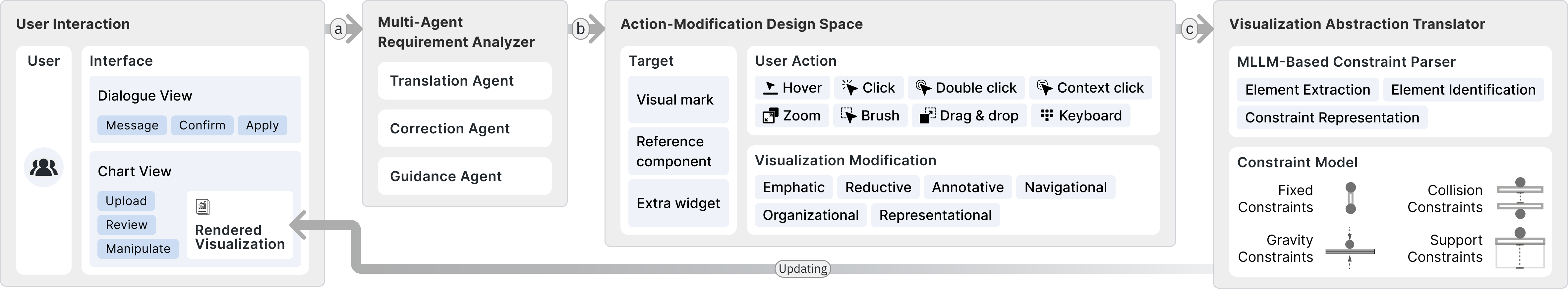}
    \caption{
    Users can upload existing visualizations in the chart view and present their authoring requirements (a) in the dialogue view.
    The multi-agent requirement analyzer employs translation, correction, and guidance agents that work collaboratively to translate users' requirements into specifications defined in (b) is the action-modification design space.
    The uploaded visualizations are abstracted to constraint-based presentations, and the analyzed modifications are also represented through constraints (c).
    Driven by these constraints, the visualization becomes interactive.
    }
    \label{fig:pipeline}
\end{figure*}

\textbf{NLI for existing visualizations.}
While the approaches mentioned above focus on constructing visualizations from data, some methods~\cite{setlur2016eviza, hoque2017evizeon, chartQA, Lai2020Annotation, kahou2017figureqa, vaithilingam2024dynavis} focused on existing visualizations.
Eviza~\cite{setlur2016eviza} and Evizeon~\cite{hoque2017evizeon} convert natural language input into filters applied to visualizations.
Kim et al.~\cite{chartQA} generate explanations to answer questions related to existing visualization charts.
Some methods aim to generate natural language content for existing visualizations, for example, generate description~\cite{liu2020autocaption} and title~\cite{liu2023autotitle}.
Instead of producing explanations, Lai et al.~\cite{Lai2020Annotation} emphasize highlighting charts to help users better understand them.

While previous methods mainly directly translate users' natural language input into specifications, recent research focuses on enhancing accuracy through iterative feedback.
We propose a multi-agent analyzer that not only translates users' input, but also corrects LLM output and provides guidance for user interaction.

\subsection{Visualization Enhancement}

\revision{Several approaches aim to enhance visualizations.
Interaction+\cite{lu2017interaction} enhanced web visualizations by parsing visual mark attributes and developing an interaction add-on, focusing on non-spatial attributes like color and opacity.
Building on D3 specifications, Harper and Agrawala\cite{harper2014deconstructing, harper2018converting} developed tools for deconstructing D3 visualizations, enabling reuse with Vega-lite templates\cite{satyanarayan2017vegalite}.
Beyond D3 visualizations, DataWink \cite{xie2025datawink} employed LLMs to enable the reuse of general SVG visualizations.
VisDock\cite{choi2015visdock} provided a system allowing programmers to add interactions (selecting, filtering, navigation) to existing visualizations through code.
PapARvis~\cite{zhu2020paparvis} and VisAR~\cite{kim2017visar} focus on enhancing visualization in augmented reality environments.
Liu et al.~\cite{liu2024spatial} proposed a spatial constraint-based method for modeling the spatial relationships between visualization elements.
Similarly, DIVI~\cite{snyder2023divi} introduced a framework for adding interactions to existing charts.
However, DIVI lacked awareness of spatial layout constraints. For example, if a user deletes a lower segment in a stacked area chart, DIVI cannot automatically shift the remaining elements to maintain visual integrity.
MoVer~\cite{MaAgrawala2025MoVer} introduced a formal verification language to check spatio-temporal properties of generated animations and feeds errors back to the LLM for correction.
In contrast, our work empowers users to define customized interactions according to their specific needs, thereby offering greater flexibility across a diverse range of scenarios and use cases.}

%% file: chapter/3_overview.tex
\section{Overview}

\toolname{} enables users to add interactions to existing visualizations based on natural language requirements.
We address three key challenges: structuring interaction possibilities, interpreting user requirements, and supporting diverse visualization implementations.
As shown in \autoref{fig:pipeline}, \toolname{} consists of three parts:
\begin{itemize}[noitemsep,topsep=2pt,parsep=0pt,partopsep=0pt]
\item An \textbf{action-modification interaction design space} that abstracts interactions into \textit{user actions} and \textit{visualization modifications} (\autoref{sec:interaction_space}).
\item A \textbf{multi-agent requirement analyzer} that analyzes users' natural language instructions into actionable interaction specifications (presented by actions and modifications) (\autoref{sec:user_input}).
\item A \textbf{visualization abstraction translator} that converts existing visualizations into a flexible representation (\autoref{sec:constraint_representation}), independent of the underlying toolkits or platforms.
This representation establishes a foundation for visualization modifications.
\end{itemize}

%% file: chapter/4_actionspace.tex
\section{Action-Modification Design Space}
\label{sec:interaction_space}





Interaction in a visualization system is a dynamic process where users manipulate data or its representation to explore, analyze, or derive insights~\cite{card1999readings}.
Drawing from foundational works in visualization research~\cite{heer2012interactive, yi2007toward}, we summarize interactions in visualization into two key components: user actions and visualization modifications.
A \textbf{modification} is defined as the response or outcome that occurs after a user performs an \textbf{action}, and a \textbf{target} is the object of an action or a modification.
We discuss the design space of target, action, and modification in the following subsections.


\subsection{Target}


A target refers to the object of action and modification.
For example, when users hover over a bar, the bar constitutes the target of this action.
Targets can be categorized into three types:

\begin{itemize}[noitemsep,topsep=2pt,parsep=0pt,partopsep=0pt]
\item \textbf{Visual mark} is the visual element that represents data in visualizations.
Visual marks include forms such as \textbf{data points, lines, bars, and areas}, which convey information through visual attributes such as \textbf{position, size, and color}.
For instance, in a bar chart, each \textbf{bar} is a visual mark, with its height corresponding to a specific value. Similarly, the lines of a line chart are visual marks.
\item \textbf{Reference component} refers to visualization components that are not directly bound with data, e.g., coordinate axes and legends.
The axes can encode categorical, temporal, and quantitative data attributes.
The legend can encode data attributes to color.
For example, in a bar chart, the $x$ axis represents categorical labels, while the $y$ axis denotes the quantity of each category.
\item \textbf{Extra widgets} encompass additional UI elements that facilitate interaction with the visualization, such as \textbf{buttons}.
For instance, users may want to click an extra button to reorder the bars in a bar chart based on their magnitude.
\end{itemize}


\subsection{User Action}



User actions are the means by which users express intentions through input devices (mouse, touchpad, or keyboard) to manipulate targets such as visual marks, reference elements, or widgets.
Users often begin by examining visual objects without altering the display. For example, they may \textit{hover} over elements to view details.
Once specific data items attract attention, users can select points of interest, either by \textit{clicking} a single element (e.g., a data point in a scatter plot) or \textit{brushing} a group of elements within a region.
After selection, users may rearrange elements using \textit{drag and drop}, or adjust the view with \textit{zoom} or \textit{double click} to explore data at different scales.
Some actions have fixed semantic meanings, such as pressing ``delete'' to remove a selected element or performing a context click to access additional options.
Building on intentions identified in prior frameworks, our eight-category scheme restructures and clarifies existing taxonomies whose boundaries and action-level mappings remain vague for computational use.

\begin{itemize}[noitemsep,topsep=2pt,parsep=0pt,partopsep=0pt]

\item \wsicon{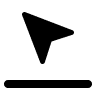} \textbf{Hover} is the action of hovering the cursor over a specific element without clicking.
For example, in a line chart, hovering over a data point may display a tooltip with additional information.

\item \wsicon{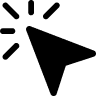} \textbf{Click} refers to a single, discrete action performed via an input device (e.g., mouse, touchscreen) to select or activate a specific element in a visualization.

\item \wsicon{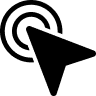} \textbf{Double-click} refers to two rapid consecutive presses or taps on the same element using an input device (e.g., mouse or touchscreen).  
It is often used to trigger a zooming response in visualizations.

\item \wsicon{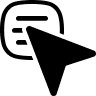} \textbf{Context-click} refers to an access input operation via an input device (e.g., mouse, touch screen) on a specific element, typically to access a context-sensitive menu or additional options.
It can be a click using a mouse, a double tap on a touch screen, or a long-press on a touch screen.

\item \wsicon{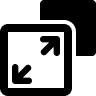} \textbf{Zoom} is the action of enlarging or reducing a specific area, often performed using the mouse wheel or pinch gestures on a touchpad.
For instance, zooming on the X-axis by scrolling the mouse wheel or pinching on a touchpad can adjust the visible data range, allowing users to focus on specific sections of the dataset.

\item \wsicon{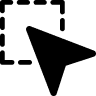} \textbf{Brush \& Drag} refers to the action of selecting a range or region by clicking and dragging the mouse over an area.
For example, in a scatter plot, brushing can be used to select a subset of data points for further analysis.

\item \wsicon{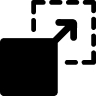} \textbf{Drag \& drop} involves clicking on an element, moving the mouse while holding the button, and then releasing it.
For example, in a \textbf{bar chart}, dragging a bar in a stacked bar chart can change the order of the stacked elements. 

\item \wsicon{Keyboard} \textbf{Keyboard} is the action of pressing the keys.
For example, users can press the ``delete'' key to remove selected visual elements.
\end{itemize}

\subsection{Visualization Modification}
\label{sec:modifications}



\begin{figure}[!ht]
    \centering
    \includegraphics[width=\columnwidth]{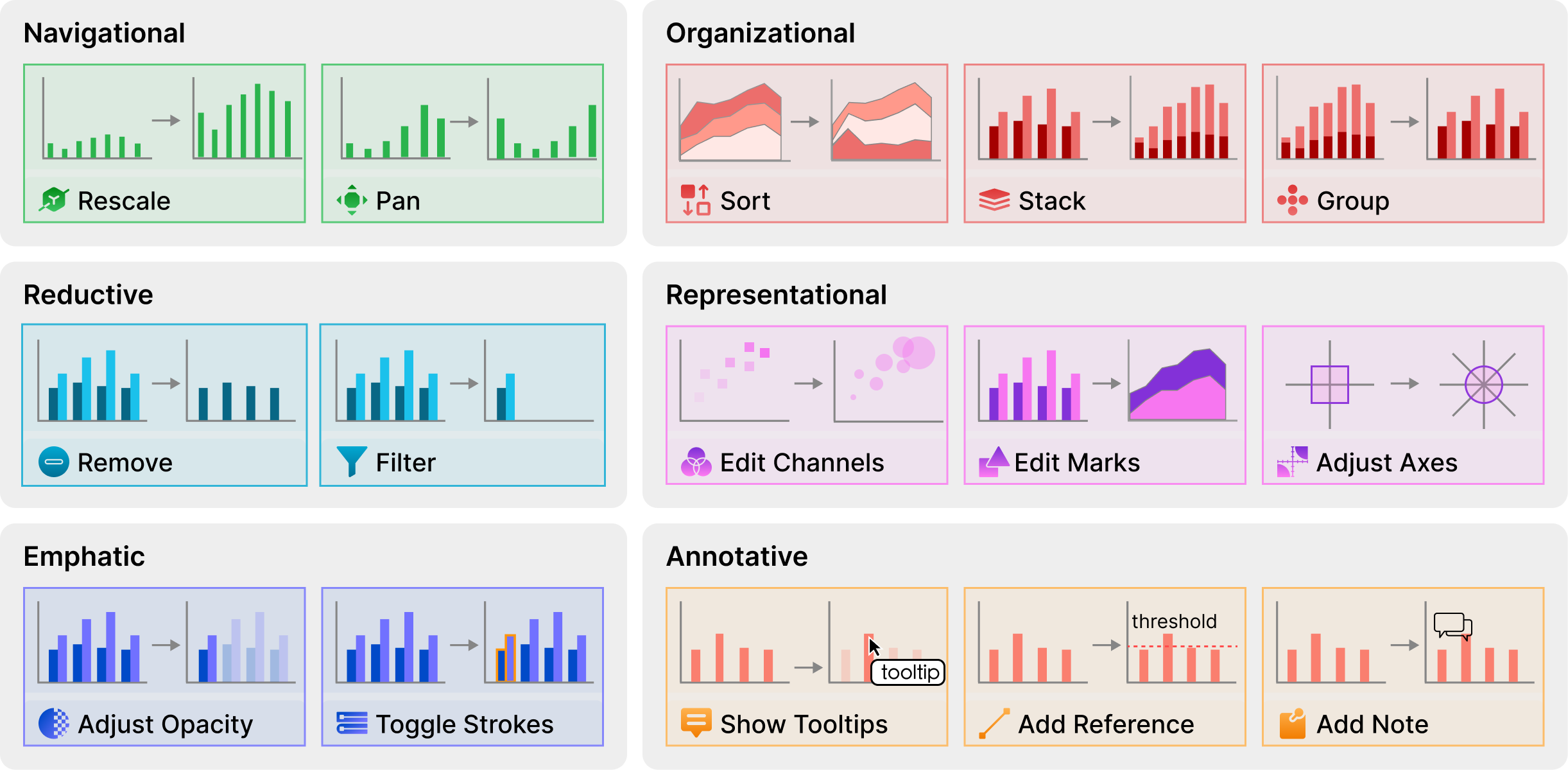}
    \caption{The design space of visualization modifications. }
    \label{fig:visoperation}
\end{figure}

Modification is defined as the response of the visualization after users perform actions.
In the field of visualization interaction, the modifications have been classified in different ways~\cite{yi2007toward, munzner2014visualization, heer2012interactive}.
Building upon these approaches, we summarized visualization modifications on a single existing chart based on the extent of changes made to the elements.
Our classification considers factors including whether the elements' positions are changed, whether elements are added or removed, and whether the current encoding method is maintained.
As depicted in \autoref{fig:visoperation}, we categorized visualization operations into six types.


    
    
    

\begin{itemize}[noitemsep,topsep=2pt,parsep=0pt,partopsep=0pt]
    \item \textbf{Emphatic} modification involves assigning varying visual intensities (e.g., opacity) to existing elements, distinguishing between user-focused and non-focused content.
    In our scenario, we have two kinds of emphatic modifications, namely, opacity manipulation \wsicon{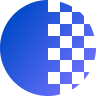} and stroke emphasis \wsicon{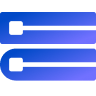}.
    Opacity manipulation lowers the opacity of unfocused elements to emphasize focused ones.
    Border emphasis adds outline to focused elements.
    This modification relates to the ``select'' operation in both Yi et al.~\cite{yi2007toward} and Munzner's~\cite{munzner2014visualization} classifications.
    
    \item \textbf{Reductive} modification selectively retains focused elements while eliminating unfocused ones.
    Similar to highlighting in its intent to prioritize user-focused content, the reduce operation goes further by completely removing non-essential elements.
    We support two kinds of reductive modification, direct removing unfocused marks \wsicon{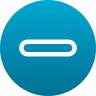} and removing according to filter condition \wsicon{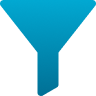}.
    Reductive modification results in a simplified visualization that concentrates on focused data.
    Reductive modification corresponds to ``filter'' in both Yi et al.~\cite{yi2007toward} and Munzner's~\cite{munzner2014visualization} classifications.
    
    \item \textbf{Annotative} modification adds elements to current visualizations to give explanations on some visual marks.
    We support three types of annotative modifications: (1) tooltips \wsicon{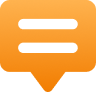} (e.g., appear on hover), (2) reference lines \wsicon{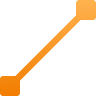} (e.g., highlight thresholds or aggregates), and (3) text labels \wsicon{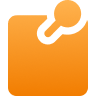} that directly annotate data points.
    This modification corresponds to ``annotate'' in Munzner's~\cite{munzner2014visualization} classification.
    
    \item \textbf{Navigational} modification alters the view for visual marks.
    The modification in our approach includes two main operations: scaling \wsicon{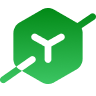} and panning \wsicon{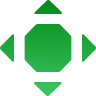}.
    Scaling changes the magnification level of the visualization (e.g., zooming into a selected area of a scatterplot).
    Panning adjusts which portion of the visualization is visible in the viewport.
    Both operations can be implemented by changing the domain and range of the axes.
    This modification does not alter the data itself but reframes the view, ensuring that the displayed content is appropriately scaled and positioned.
    Navigational modification belongs to ``reconfigure'' in Yi et al.'s~\cite{yi2007toward} framework and corresponds to ``navigate'' in Munzner's~\cite{munzner2014visualization} classification.
    
    \item \textbf{Organizational} modification involves altering the arrangement of elements within a visualization. Reorganization techniques include sorting \wsicon{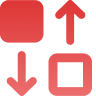}, stacking \wsicon{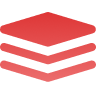}, overlapping \wsicon{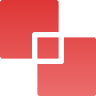}, and grouping \wsicon{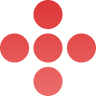}.
    For instance, stacking can aggregate data into a summarized form (e.g., transforming grouped bars into stacked bars), while sorting elements along an axis can enhance ranking or comparison tasks.
    The modification belongs to ``reconfigure'' in Yi et al.'s~\cite{yi2007toward} framework and ``rearrange'' in Munzner's~\cite{munzner2014visualization} classification.
    
    \item \textbf{Representational} modification changes the visual encoding of a chart.
    We consider three types:
    (1) changing visual channels \wsicon{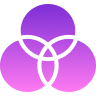} (e.g., color, size, or shape);
    (2) \revision{changing mark types} \wsicon{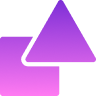} (e.g., converting a line chart to an area or bar chart); and
    (3) changing axes \wsicon{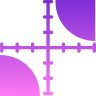} (e.g., switching from a linear to a logarithmic scale).
    Changing the representation type is essential when the current encoding cannot support a required analysis. For example, in a multi-line chart of product sales trends, if the user wants the overall trend, re-encoding transforms it into a stacked area chart, where each mark represents a product and the total height shows the sum.
    \textit{Representational} corresponds to ``encode'' in Yi et al.~\cite{yi2007toward} and ``change'' in Munzner~\cite{munzner2014visualization}.
\end{itemize}

%% file: chapter/5_nlinput.tex
\section{Multi-agent Requirement Analyzer}
\label{sec:user_input}

The goal of the multi-agent requirement analyzer is to parse the user's requirements into interaction specifications (actions and modifications as defined in \autoref{sec:interaction_space}).

\subsection{Requirements}

Parsing users' natural language instructions presents three main challenges: \textbf{Interpretation:} Users often lack technical terminology and describe the same interaction in varied ways (e.g., ``mouseover'' vs. ``hovering''). The system must map such diverse expressions to precise specifications.
\textbf{Execution:} Even well-interpreted instructions may yield technically infeasible specifications. The system must validate and refine results to ensure executability.
\revision{\textbf{Specificity:} Users frequently give ambiguous and under-specified instructions (e.g., ``make this chart interactive'')~\cite{setlur2022you}.}
The system must detect missing details and guide users toward complete specifications.

\begin{figure}[htbp]
    \centering
    \includegraphics[width=.8\columnwidth]{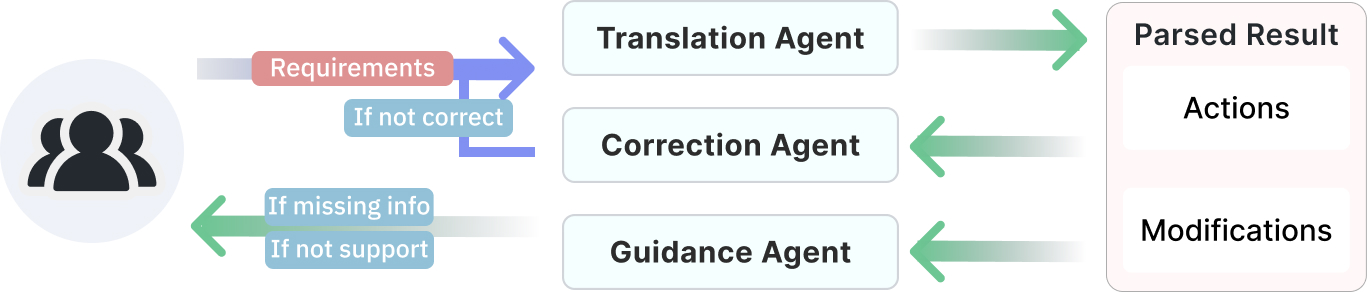}
    \caption{The process of LLM parsing users natural language input.}
    \label{fig:multiagent}
\end{figure}

To address these challenges, we introduce a multi-agent architecture (see \autoref{fig:multiagent}), consisting of a \textit{translation agent}, a \textit{correction agent}, and a \textit{guidance agent}.
This multi-agent analyzer provides a robust system for translating natural language requirements into formal interaction specifications.
The agents collaboratively guide users in authoring interactions, highlight errors in their instructions, and refine LLM-generated results, ensuring the final design meets explicit needs, addresses implicit expectations, and aligns with visualization interaction principles.


\subsection{Translation Agent}
This agent transforms user instructions into formal interaction specifications.
Users often express their needs in natural language that can be ambiguous or underspecified.
The translation agent interprets these instructions and maps them to our action-modification interaction model in \autoref{sec:interaction_space}, ensuring that user intent is accurately reflected in the formal representation.
The agent takes the user requirements, supported actions, and modifications as selectable inputs and outputs a structured interaction specification with determined user actions and modifications.
Each action and modification has a target.
When the actions happen, the modifications will be triggered.
For example, the translation agent processes the instruction \textit{``Create a button that, when clicked, changes the chart to a bar chart.''} as follows.
The term \textit{``click''} is directly mapped to the action click~\wsicon{Click}.
The action target is an extra widget, a button.
The modification ``change'' is mapped to representational modification, and the parameter is ``bar chart''.
The system will create a button, when users click the button, the system will reencode the visualization into a bar chart.








\subsection{Correction Agent}

The correction agent acts as a validator, refining the translation agent's output to ensure accuracy and alignment with the user's intent.
The correction agent reviews the output from the translation agent, identifying inconsistencies or technical infeasibilities that may arise from misinterpretations or limitations of multimodal large language models (MLLMs), and provides feedback to the translation agent.
The correction agent checks the following:

\begin{itemize}[noitemsep,topsep=2pt,parsep=0pt,partopsep=0pt]
\item Check if the parsed results follow the action-modification structure required.
The output should be a list of actions and modifications. For each action and modification, it verifies whether the target is compatible with it.
\item Check if the users' requirements are satisfied using the output action and modifications. It confirms whether the translation agent's output accurately reflects the user's instructions.
\item Check if the output modifications can be supported in the visualization: For example, stacking or grouping is not supported in a line chart, and the correction agent validates this.
\item Check if additional modifications are needed: For instance, after removing some visual marks, the correction agent assesses whether rescaling the axes is necessary to optimize space usage.
\end{itemize}

By cross-referencing the translation agent's output with the design space and the user's original instructions, the correction agent enhances the system's reliability, making it more accessible to non-expert users. If discrepancies are detected, the correction agent provides feedback and requests the translation agent to regenerate the parsed results.






\subsection{Guidance Agent}\label{sec:guideagent}

The agent identifies gaps in user requirements, highlights errors or ambiguities, and suggests complementary interactions.
Because users often have incomplete mental models of possible or useful interactions, the guidance agent refines specifications by recommending contextually relevant interactions and steering users toward clearer, more effective requests.
\textbf{Missing required information}.
If users' instructions do not provide enough information about actions and modifications, the guidance agent will point out the missing information and give some feedback.
If the possible action, modification have a default choice that is most commonly chosen, the guidance agent would produce the possible choice and ask to confirm.
For example, a user inputs ``show tooltip''.
Since no action and action target are explicitly mentioned, the guidance agent assumes the most common action, mouse over a visual element.
The guidance agent assigns \textit{visual mark} as the target of the action and \textit{hovering} as the action.
\textbf{Out of range.}
When users' requirements are out of the current support range, the guidance agent will give feedback that points out the current action or modifications that are not supported and ask users to provide other requirements.







%% file: chapter/6_representchart.tex
\section{Visualization Abstraction Translator}
\label{sec:constraint_representation}


This section presents our approach to abstracting visualizations into an implementation-agnostic representation: we first discuss the rationale for this representation (\autoref{sec:consideration_representation}), then introduce an MLLM-based parser for converting visualizations (\autoref{sec:constraint_parse}), and finally describe how the model supports visualization modifications (\autoref{sec:constraint_modification}).


\subsection{Consideration for Representation}
\label{sec:consideration_representation}

We require an implementation-agnostic representation for static visualizations that supports both cross-platform adaptability and subsequent modifications. The core requirements are to preserve the appearance of existing visualizations while enabling modification. Existing approaches range from high-level specifications such as Vega-Lite~\cite{satyanarayan2017vegalite} to low-level implementations such as D3~\cite{bostock2011d3}. Low-level representations precisely preserve appearance by directly controlling visual elements (for example, shapes, lines, and positions), but they lack structured abstractions that connect elements to data semantics, which makes modification difficult. High-level specifications, in contrast, encode abstract structures such as categories, axes, and data bindings. They are often used as intermediate representations in prior work~\cite{harper2014deconstructing, harper2018converting}. However, when high-level specifications are used to recreate an existing visualization, they regenerate all marks from scratch rather than reusing the original ones. As a result, the new marks no longer correspond directly to the originals, making it difficult to maintain visual consistency during subsequent interactions or modifications. Neither purely high-level specifications nor low-level implementations fully meet our needs. We therefore seek an intermediate representation that captures the appearance of existing visualizations while maintaining the flexibility required for modification.

\begin{figure}[ht]
    \centering
    \includegraphics[width=.85\linewidth]{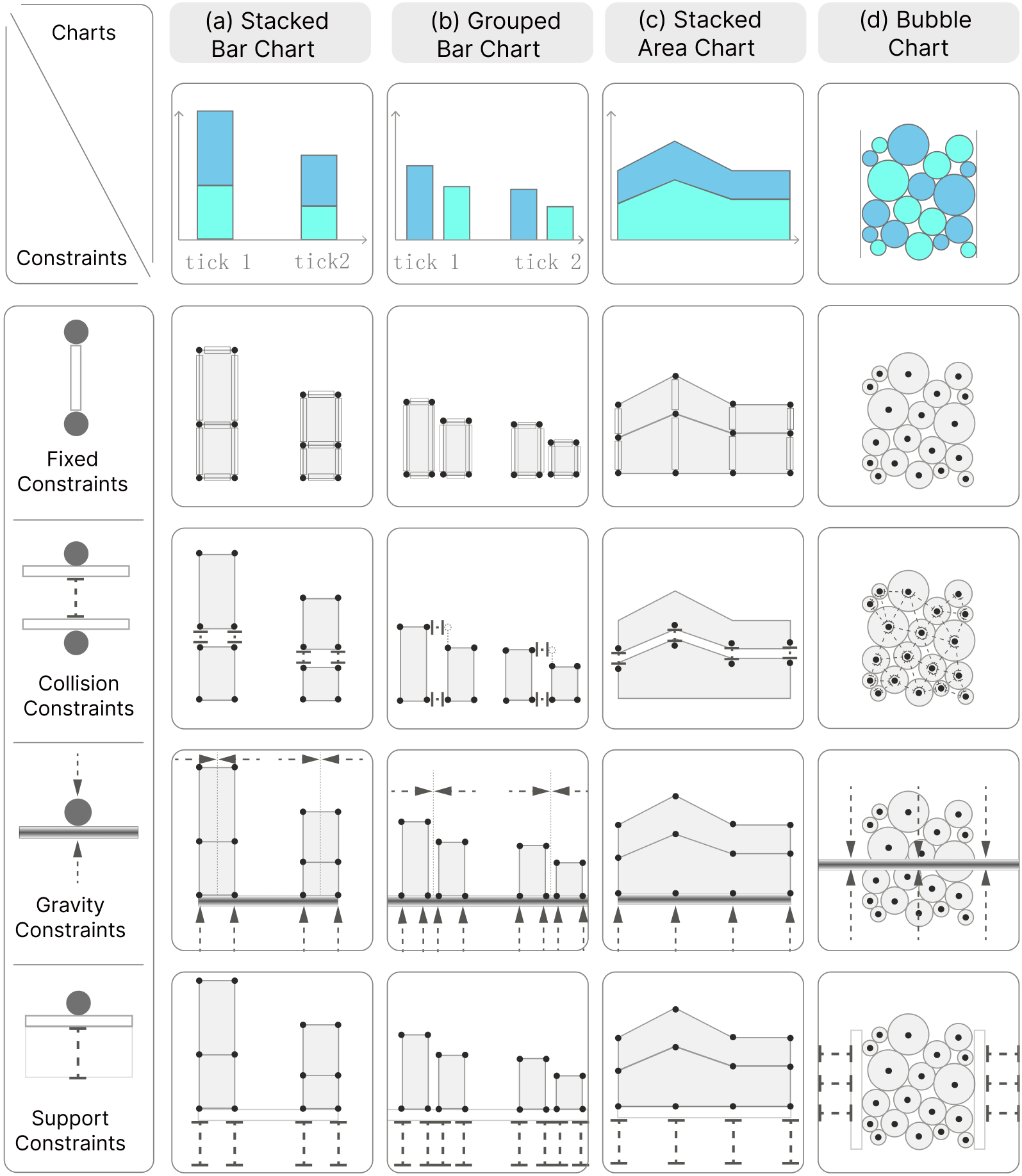}
    \caption{The constraint model~\cite{liu2024spatial} is employed as the representation.}
    \label{fig:constraint_model}
\end{figure}

Thus, we use a constraint-based representation~\cite{liu2024spatial} that abstracts visualizations modeled by control points and constraints.
This representation meets our needs by \textbf{preserving appearance} while \textbf{enabling smooth, automatic modifications} through constraints.
It fully maintains the appearance of original visual marks (e.g., shapes, colors, and sizes) and uses control points to represent different visual marks, such as employing contour lines to characterize polygons and paths. It expresses data mappings and spatial dependencies through various types of spatial constraint relationships.
The constraints can drive visual marks to finish modifications fluently and automatically.
These spatial constraints can be transformed into forces that guide the positions and positional changes of control points, allowing modifications to existing visualizations.
For example, in a stacked bar chart where each bar represents a category, if one category is removed, the remaining bars automatically shift downward to fill the gap, guided by spatial constraints.
In the constraint model, the \textbf{visual marks} (points, lines, areas, and text) are presented by \textbf{control points} and their connections, as depicted in \autoref{fig:constraint_model}.
The constraint model formulates the spatial positions of visual marks using three types of relationships.





\begin{itemize}[noitemsep,topsep=2pt,parsep=0pt,partopsep=0pt]

\item \textbf{Intra-mark relationships.}
The constraint among control points within a visual mark defines the intrinsic shape and size.
For instance, the width and height of a bar can be determined by adding \textit{fixed constraints} in the corners (\autoref{fig:constraint_model} (a) and (b)).

\item \textbf{Inter-mark relationships.}
The constraints can describe the relative positions among visual marks.
For example, the area visual mark in a stacked bar chart (\autoref{fig:constraint_model} (a)) has vertical \textit{collision constraints}, which force a visual mark on top of another.

\item \textbf{Mark-to-external relationships.}
Constraints further express the relationships between a visual mark and external references.
For example, an area mark in a stacked area chart is aligned with the $x$-axis, where the visual mark is driven by the \textit{support constraint} and \textit{gravity constraint} from the $x$-axis (\autoref{fig:constraint_model} (c)).

\end{itemize}

\subsection{MLLM-based Constraint Parser}
\label{sec:constraint_parse}
We propose an algorithm that leverages an MLLM to parse existing visualizations into a constraint-based representation.
The parser operates in three stages: (1) extracting visual objects from the chart, (2) identifying the chart type and the roles of these objects, and (3) constructing constraints based on the identified structure.


\subsubsection{Element Extraction}
We employ JavaScript-based parsing scripts executed in a headless browser to analyze SVG files and extract visual elements. These elements are converted into a simplified vector format (SimVec~\cite{liu2025simvec}) for efficient processing.
We identify SVG elements such as \texttt{<rect>}, \texttt{<text>}, and \texttt{<path>}.
\revision{During this process, all CSS classes and inline styles are resolved into their final rendered properties, such as fill color, opacity, stroke width, and stroke color, while redundant style metadata and class identifiers are discarded. To eliminate the complexity of nested transformations, all coordinates are converted into absolute values.}
We then convert the data into the SimVec format. This preserves essential chart information while reducing token usage by up to 90\%, in practice.
\revision{For instance, any closed path or geometric shape (e.g., path, rect, or circle) is abstracted into a structured format.
A representative example of a polygon in SimVec format is: $\text{polygon: points}[(x_1, y_1), (x_2, y_2), \dots], \text{color: } (h, s, l)$.}




\subsubsection{Element Identification}

We employ an MLLM to classify the role of elements in a chart, for example, the axis tick, the title, and the legend.
The SimVec data, combined with the corresponding bitmap representation of the SVG, are provided as inputs to the large language model.

\begin{itemize}[noitemsep,topsep=2pt,parsep=0pt,partopsep=0pt]
    \item \textbf{Element-level identification}:
    We employ an MLLM to classify the role of the elements, which is also the fundamental step of previous reverse engineering work for visualization~\cite{poco2017rev, savva2011revision}.
    Each element in a chart may be a data-encoded element or reference component.
    The data-encoded elements are those visual marks that directly encode data, e.g., a bar in a bar chart or a circle in a scatter plot.
    The reference elements are those elements that provide extra information for the data elements, for example, the legend or the title of a chart.
    \item \textbf{Chart-level identification}: 
    The chart-level attributes are inferred by the MLLM, e.g., the chart type, axes type, and whether there are stacking/grouping/overlapping relationships.
\end{itemize}

\subsubsection{Constraint Representation}

After classifying the roles of each element, we can represent the visual marks using control points and constraints.
With identified reference elements (legends, axes) and data-encoded elements (e.g., bars, points), we establish the complete data mappings for the input chart.
We determine the precise spatial coordinates of each data element and their corresponding data values by calculating the axes scales and legend mappings based on the element classifications.
This allows us to recover both the visual properties (position, size, color) and the corresponding data values for each mark in the visualization, effectively reconstructing the data-to-visual encoding relationships. 

We organize constraints into steps (see \autoref{fig:constraint_model}).
We add \textit{fixed constraints} to visual marks. In bar charts, we establish x and y fixed constraints between vertices of connected edges. For area marks, we apply fixed constraints between upper and lower points sharing the same x-coordinate.
We apply \textit{collision constraints} between control points of stacked elements. In stacked area charts, collision constraints prevent overlap between stacked marks.
We apply \textit{support constraints} to points of marks stacked on axes, ensuring they remain above the x-axis.
\textit{Gravity constraints} have two roles: marks stacked on axes experience forces toward the axes, ensuring alignment; in directions without stacking, constraints pull marks toward data value ticks. For example, in line charts, each point is pulled toward its data value. Gravity constraints ensure that elements respond to axis scale transformations.


\subsection{Constraint-Based Modification}
\label{sec:constraint_modification}

We translate modifications in \autoref{sec:modifications} into constraint changes.
\begin{itemize}[noitemsep,topsep=2pt,parsep=0pt,partopsep=0pt]
 \item \textbf{Emphatic} modification doesn't alter the position or structure of the control points, and no constraint adjustments are needed. Emphatic modification is achieved by changing the opacity~\wsicon{opacity} or stroke~\wsicon{stroke} of selected and unselected visual marks.
 \item \textbf{Reductive} modification removes some visual marks from the visualization.
 When a mark is removed, the constraints automatically drive the positions of the remaining visual marks.
 For example, in a stacked area chart, when the bottom visual mark is removed, the upper ones automatically drop down.
 \item \textbf{Annotative} modification does not affect the spatial layout of visual marks, no constraint changes are needed. We directly add tooltips \wsicon{tooltip}, lines \wsicon{line}, and text labels \wsicon{notes} on top of the visualization.
 \item \textbf{Navigational} modification (i.e., rescaling \wsicon{rescale_axes} and panning \wsicon{panning}) involves changing the axis scale (e.g., zooming in on a bar chart).
 In the constraint model, constraints of control points are tied to the relative axes.
 When the scale of the axis changes, the related fixed, gravity, and support constraints are updated.
 \item \textbf{Structural} modification includes sorting, stacking, overlapping, and grouping.
 For sorting \wsicon{sorting} on axes, the control points representing these marks are governed by constraints tied to the axis (e.g., order bars by height).
 For stacking \wsicon{stacking}, grouping \wsicon{grouping}, and sorting of stacking \wsicon{sorting}, the order or direction of collision constraints is altered.
 For example, changing from a stacked bar chart to a grouped bar chart changes the collision constraint order from horizontal to vertical.
 \item \textbf{Representational} modification changes the visual mark types.
 For visual channel change~\wsicon{visual_channel} and representation type change ~\wsicon{representation_type}, we calculate new control points from existing ones to preserve spatial consistency.
 Constraints among new control points are given accordingly.
 For example, when we change a line chart to a bar chart, each point in the line chart is replaced by a new visual mark with four control points and fixed constraints among these points.
 For axes change~\wsicon{change_axis}, we change the coordinate type or scale of axes and update related constraints accordingly.
 
\end{itemize}

\section{Interface}

\begin{figure}[ht]
    \centering
    \includegraphics[width=\columnwidth]{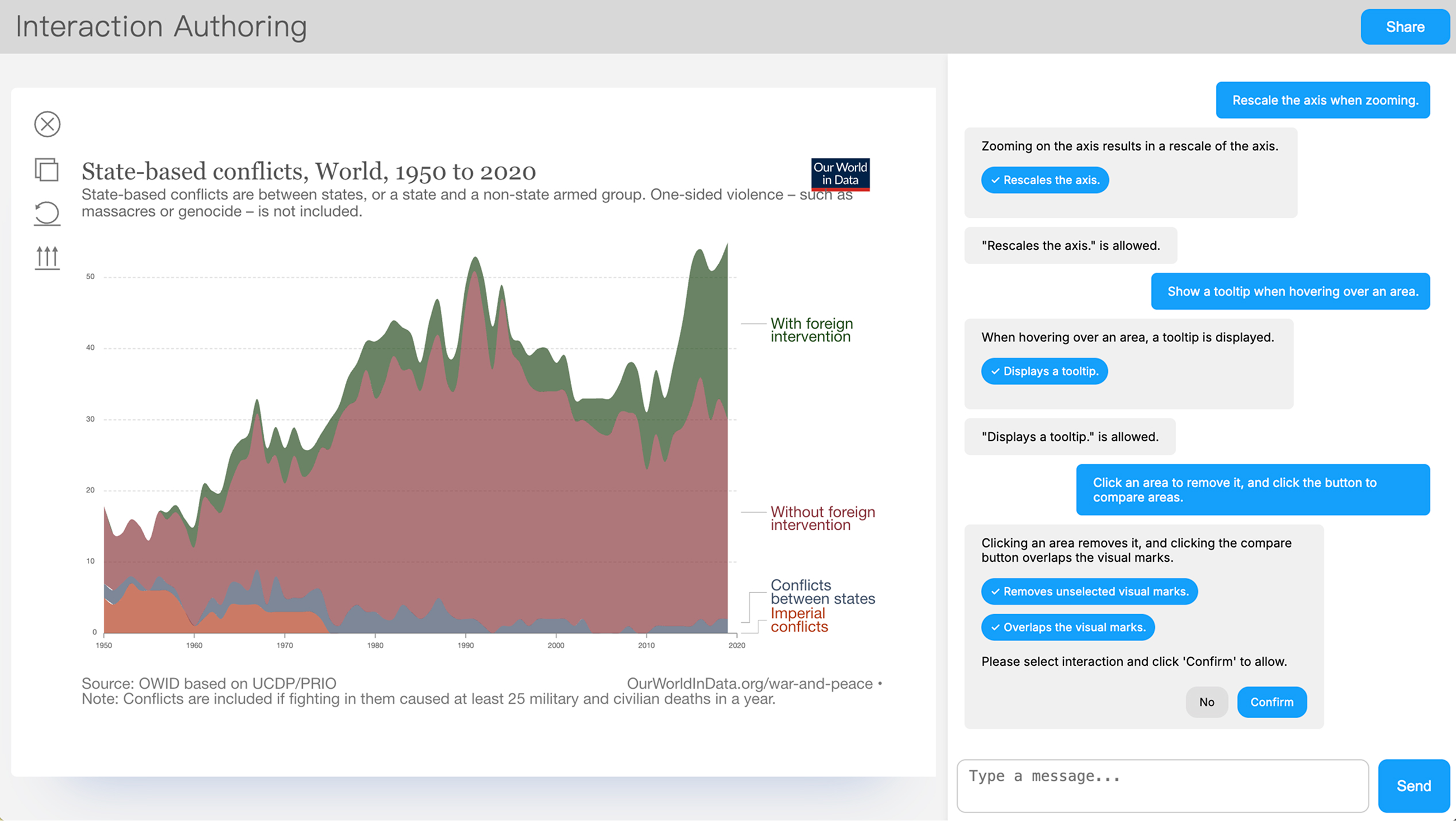}
    \caption{The interface of the visualization interaction authoring system. On the left is the chart view; on the right is the dialogue view.}
    \label{fig:interface}
\end{figure}

As shown in \autoref{fig:interface}, the interface has two views: a dialogue and a chart view.  
The creator can upload an SVG in the chart view, and specify their authoring requirements in the dialogue view.



\textbf{Dialogue view.}The dialogue view enables creators to author visualization interactions using natural language.
Users can specify visualization interactions in natural language, and the multi-agent analyzer (\autoref{sec:user_input}) translates these into structured interaction specifications.
The system also provides guidance through the dialogue view when creators provide under-specified or out-of-range instructions.
The dialogue process consists of three stages:

\begin{itemize}[noitemsep,topsep=2pt,parsep=0pt,partopsep=0pt]
\item \textbf{Message}: Users enter authoring instructions in natural language through the interactive interface. For example, a user might type, \textit{``I want to rescale the x-axis when I zoom on the axis.''}

\item \textbf{Confirm}: After the user inputs an instruction, the analyzer converts it into a structured action-modification specification.
The system then displays an NL explanation of this specification, detailing the actions and modifications to be applied to the uploaded visualization. Creators can refine the modifications by deselecting unwanted options. To proceed, the creator confirms the action-modification specification by clicking the ``Confirm'' button in the response.

\item \textbf{Apply}: Once the user confirms the added interaction, the system applies the corresponding actions and modifications to the visualization in the chart view.
\end{itemize}

\textbf{Chart view.}
Once a visualization is uploaded, it is immediately converted into a constraint-based representation.
This representation retains the visual appearance of the uploaded visualization but alters its underlying implementation by replacing the visual marks with control points and their connections.
After the creator adds and confirms actions and modifications in the dialogue view, these interactions are applied in the chart view.
The actions are directly translated into inputs for corresponding hardware devices, such as mouse and keyboard inputs, and are associated with event listeners for the targeted elements.
If the targets of the actions are additional widgets, the system creates these new widgets within the chart view.
Simultaneously, the modifications are converted into changes to control points and constraints as described in \autoref{sec:constraint_modification}.




%% file: chapter/7_evaluation.tex
\section{Evaluation}

We evaluate the usability and effectiveness of the method through two case studies (\autoref{sec:case_study}) and a user interview (\autoref{sec:user_interview}).
We focus our evaluation on case studies and user interviews, since to our knowledge no existing systems support adding interactivity to pre-existing visualizations without coding.

\subsection{Case Study}
\label{sec:case_study}


We present two cases to demonstrate how the proposed system interprets various types of natural language inputs and generates corresponding interactive modifications.
To clearly describe the scenarios involving the creation and use of interactions, we use \textit{creator} to refer to the person authoring the interaction and \textit{user} to refer to the person engaging with the interactive visualization.


\begin{figure}[ht]
    \centering
    \includegraphics[width=.9\columnwidth]{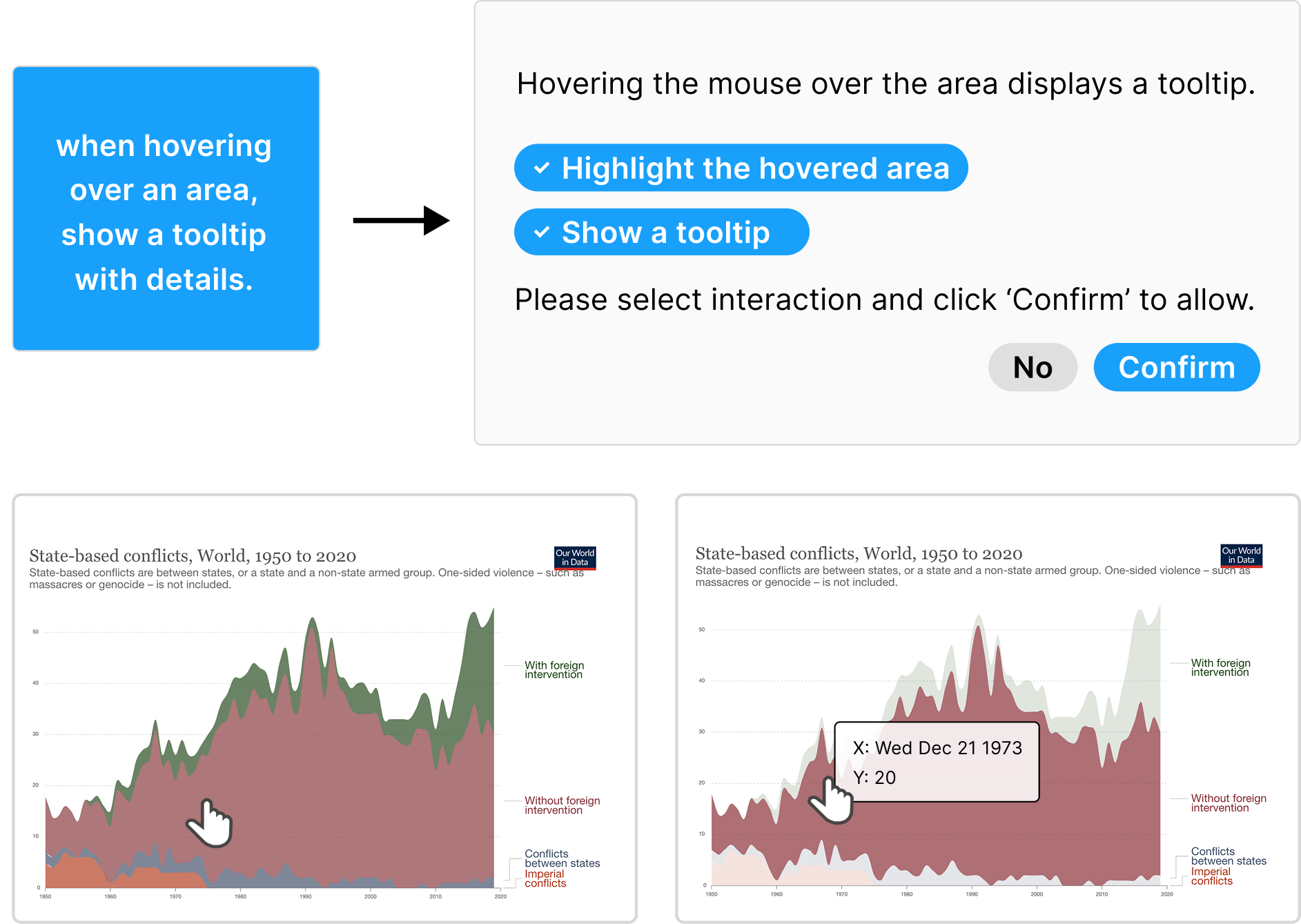}
    \caption{A creator defines an interaction: when hovering over a visual mark, a tooltip displays its details.}
    \label{fig:stacked_area_chart}
\end{figure}

\begin{figure*}[ht]
    \centering
    \includegraphics[width=.9\textwidth]{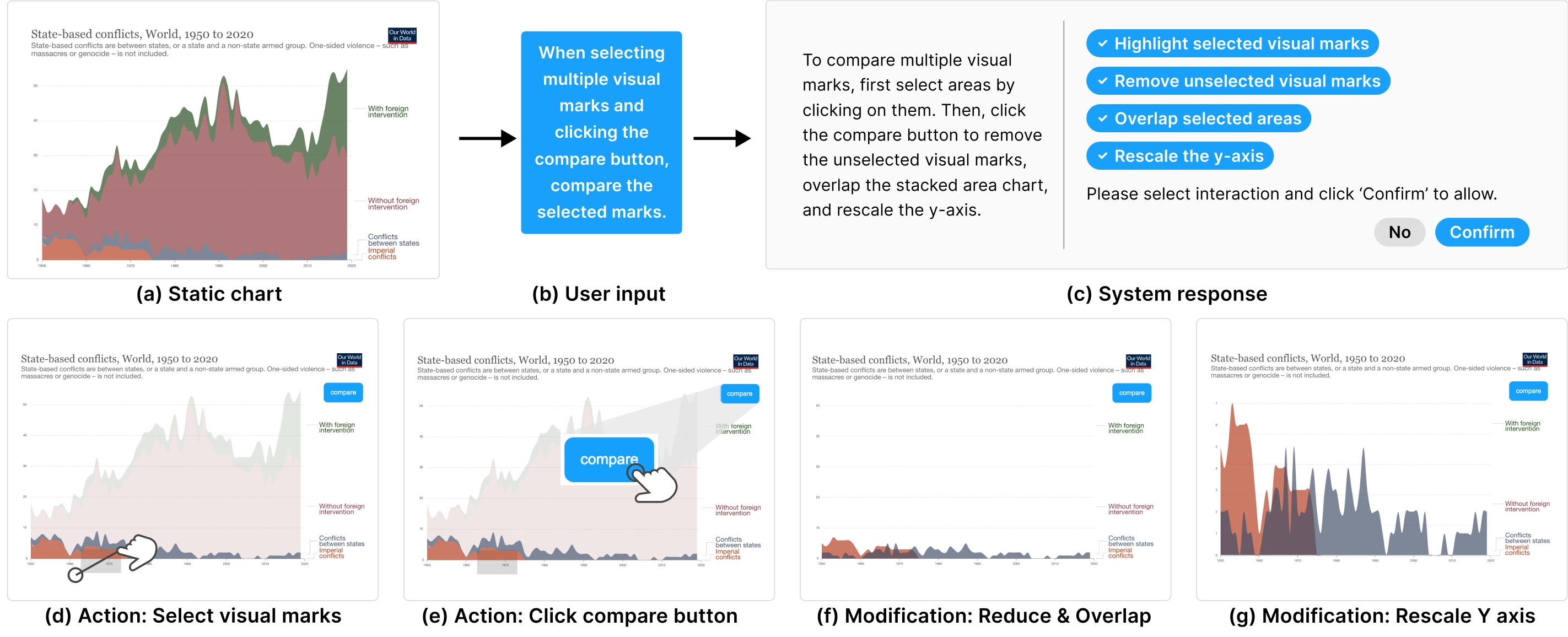}
    \caption{A creator defines a comparison interaction with two actions and four sequential modifications. After deployment, when users select multiple visual marks and click the ``compare'' button, the interactive visualization transitions from (d) to (g) to support comparison.
    }
    \label{fig:stacked_area_complex}
\end{figure*}





\subsubsection{Case 1: Augmented Area Chart}

The original visualization is a stacked area chart.
As shown in \autoref{fig:stacked_area_chart}, a creator provides the natural language input: \textit{``when hovering over an area, show a tooltip with details.''}
The system processes this input into a structured interaction specification: \texttt{\{action = ``hover'', action\_target = ``visual mark'', modification = ``annotate'', modification\_target = ``visual mark''\}}.
As depicted in \autoref{fig:stacked_area_chart}, the possible modifications derived from this specification are displayed in the dialogue view for the creator to confirm.
After adding interaction, the creator can deselect any unwanted modifications and click the confirm button.
Once confirmed, the interaction is applied to the existing chart in the chart view to create an interactive one.
After applying the interaction, when a user hovers over a visual mark (i.e., an area in this chart), the system shows a tooltip with more details about that mark.
\autoref{fig:stacked_area_complex} shows a more complex case with four modifications.
The creator authors a more complex interaction: ``When selecting multiple visual marks and clicking the compare button, compare the selected marks.''
The system parses this input into interaction specifications with multiple actions and modifications.
The actions include selecting marks and clicking an \textit{extra widget} (the compare button generated by the system).
The system suggests several modifications (\autoref{fig:stacked_area_complex} (c)), including: ``highlight selected visual marks,'' ``remove unselected visual marks,'' ``overlap selected areas (for comparison),'' and ``rescale $y$-axis (for better space usage)''.
These parsed modifications are presented as selectable boxes, and once confirmed by the creator, the system applies the corresponding actions and modifications.

After enhancing the static visualization shown in \autoref{fig:stacked_area_complex} (a) with interactive functionalities, the resulting visualization better supports users in the comparison task.
As illustrated in \autoref{fig:stacked_area_complex} (d), users can select two visual marks using a brush action.
These marks occupy only a small portion of the chart area, making their corresponding values hard to discern in the original visualization.
Moreover, because the marks are stacked, comparing them is challenging due to distortions, such as the sine illusion effect~\cite{sinestream2021, vanderplas2015signs}, where stacked areas visually exaggerate or obscure differences in magnitude.
The added interaction addresses this issue by aligning the selected visual marks, mitigating such perceptual distortions.
After selection, users click the ``Compare'' button, introduced in \autoref{fig:stacked_area_complex} (e).
In response, the system removes unselected marks, separates the stacked relationship between the selected marks (\autoref{fig:stacked_area_complex} (f)), and rescales the $y$-axis to optimize space usage (\autoref{fig:stacked_area_complex} (g)).
As a result, the selected marks are aligned and enlarged along the $y$-axis, enabling users to compare them effectively.

\subsubsection{Case 2: Exploratory emissions chart}

We use a multi-line chart from Our World in Data~\cite{owid2025sulfur} that depicts global sulfur dioxide ($\text{SO}_2$) emissions by sector (industry, energy, and transport) from 1750 to 2022.
Our system enables creators to augment such visualizations with interactions that support interactive data exploration.
For example, if a creator specifies a button for switching to an area chart, the system automatically generates the corresponding control.
Similarly, a button can be defined to switch to a bar chart.
Creators can also author interactions that allow users to drag elements within the chart to transform overlapping areas into stacked areas.
When a creator requests zooming on the $x$-axis, the system enables in/out zoom functionality to adjust the visible time range.
Additionally, creators can define tooltips that reveal detailed data values when users hover over visual marks.
Through this authoring process, the uploaded chart is transformed into an interactive visualization.

By combining different interactive functionalities, users can explore data from multiple perspectives.
As shown in \autoref{fig:bar_chart} (a), the initial view employs a line chart to illustrate emission trends from various categories over time.
When users need to understand the total contribution of each emission source, they can switch to an area chart by clicking the ``change to area chart'' button, transforming to \autoref{fig:bar_chart} (b).
To see the total trend, users can drag a visual mark up to change the overlapped areas into stacked areas (\autoref{fig:bar_chart} (c)).
Users can zoom into a specific period (e.g., 1974 - 1978) by adjusting the x-axis range (\autoref{fig:bar_chart} (d)).
However, this view may not reveal precise values.
Therefore, users can transform the visualization into a bar chart by clicking the ``change to bar chart'' button (\autoref{fig:bar_chart} (e)).
As demonstrated in \autoref{fig:bar_chart} (f), the system supports detailed data inspection.
When users hover over specific data points, the system displays an information tooltip containing precise values. For instance, the figure shows that $\text{SO}_2$ emissions from international shipping were 50.3 units in 1977.
This feature addresses limitations of visual estimation, enabling quantitative analysis, particularly in cases where multiple data series overlap or values are closely proximate.

\begin{figure*}
    \centering
    \includegraphics[width=.9\textwidth]{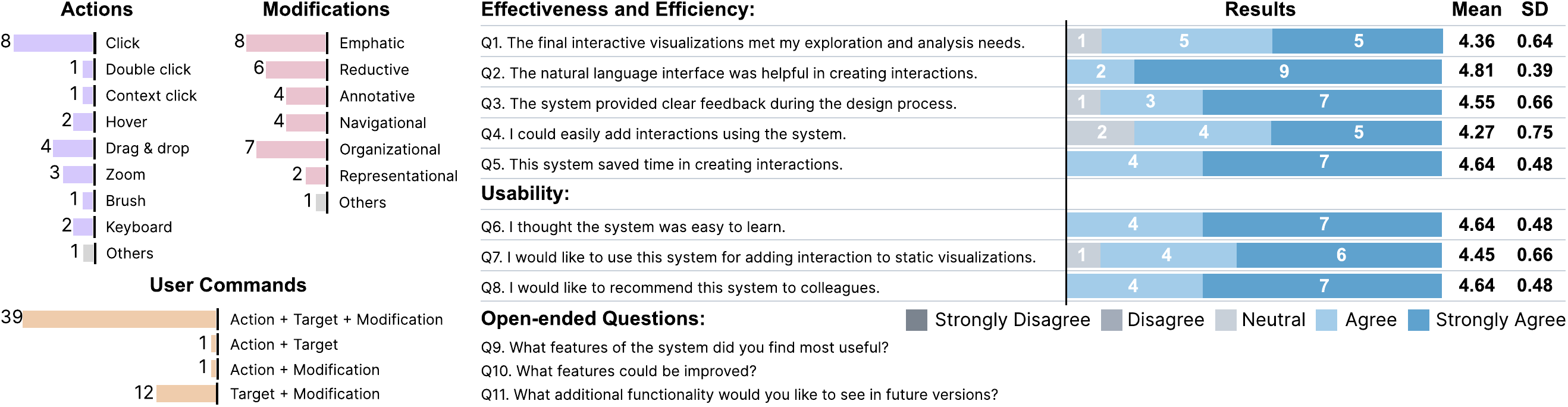}
    \caption{User interview results. Horizontal bar charts on the left show the actions, modifications, and potential commands mentioned by users during the pre-task questionnaire, indicating that our approach can cover most user needs for adding interactions to static charts. The questions in the post-task questionnaire and their results are shown on the right.}
    \label{fig:user_study}
\end{figure*}

\subsection{User Interview}
\label{sec:user_interview}

We evaluated \toolname{} through interviews with 11 participants experienced in data visualization and design.

\subsubsection{Participants and Apparatus}
\revision{
To evaluate the system's usability and generalizability across users with different backgrounds,
we recruited 11 participants (P1-P11, 5 males, 6 females) from various academic and professional fields, including mathematics, computer science, international trade, business engineering, and software development.
This cohort included two visualization experts (\textbf{P2, P6}), two data analysts (\textbf{P3, P7}), one developer (\textbf{P9}), five postgraduate students (\textbf{P1, P4, P5, P8, P11}), and one undergraduate student (\textbf{P10}). 
All participants reported prior experience in creating or analyzing visualizations as part of their academic or professional work. 
Specifically, two participants (\textbf{P3, P6}) had 3-5 years of experience, five (\textbf{P2, P4, P8, P9, P11}) had 1-3 years, and four (\textbf{P1, P5, P7, P10}) had less than one year of experience.
Our system was hosted on a server, allowing participants to access and interact with it using desktops.}


\subsubsection{Procedures}
At the beginning of each session, we introduced the background and explained our goal of transforming static visualizations into interactive ones using natural language commands.
To assess the comprehensiveness of our design space, participants first completed a pre-task questionnaire, where they listed the actions, modifications, and natural language commands they would like to use to add interactions.
Next, we conducted a 10-minute demonstration showcasing the system's core features, including how users could specify interactions through natural language and how the system would apply these interactions to static visualizations.
Following the demonstration, participants were encouraged to freely explore the system for 10 minutes to become familiar with its interface and functionality.
\revision{Then, in the task phase, participants were asked to add interactive features to static visualizations described in \autoref{sec:case_study}.}
We encouraged them to think aloud during the process, explaining the interactions they chose and their rationale for how these interactions would improve the visualization.
Participants had 30 minutes to complete this task.
Finally, participants completed a post-task questionnaire (\autoref{fig:user_study}) including eight closed-ended questions (Q1-Q8) on system effectiveness, efficiency, and usability (rated on a 5-point Likert scale~\cite{joshi2015likert}), and three open-ended questions (Q9-Q11) on useful features, improvements, and suggestions.

\subsubsection{Results}
\autoref{fig:user_study} presents the results from both the pre-task and post-task questionnaires, and the detailed analysis is as follows.

\textbf{Design Space Comprehensiveness.}
We analyzed the number (n) of actions and modifications mentioned by participants in the pre-task questionnaire and mapped them to our design space, as illustrated in the horizontal bar charts in \autoref{fig:user_study}.
Participants proposed a wide range of actions, such as \textit{Click} (n=8), \textit{Drag \& Drop} (n=4), and \textit{Zoom} (n=3), as well as corresponding modifications, including \textit{Emphatic} (n=8), \textit{Organizational} (n=7), \textit{Reductive} (n=6), and \textit{Annotative} (n=4). 
These results indicate that our system supports most user-intended actions and modifications.
A few exceptions were noted, such as an \textit{action} for adding interactive components (e.g., checkboxes) and a \textit{modification} of triggering context menus, which fall outside the current scope of our design space.
Additionally, the structure of natural language commands collected revealed a strong preference for full-form expressions in the pattern of \textit{``Action + Target + Modification''} (n=39), indicating that users naturally specify interaction intents with all three components.
Some commands only contained ``\textit{Target}'' and ``\textit{Modification}'' (n=12), often relying on implicit or default actions.  
For example, in the command ``show tooltips on the points,'' the implied action is hovering or clicking.  
\toolname{} can still support such vague commands through the aforementioned guidance agent (\autoref{sec:guideagent}).


\textbf{Effectiveness and Efficiency.}
All questions related to effectiveness and efficiency (Q1-Q5) received mean scores above 4, indicating that most participants found the system's functionality satisfactory. 
In particular, participants agreed that the system enabled them to create meaningful interactive visualizations aligned with their analytical goals (Q1) and that the natural language interface helped specify interactions (Q2).
One exception was participant P1, who gave a neutral rating for Q1. She noted the need for a toggle button to switch between two interactions that shared the same action, which the current system did not support.
Participants also generally found the system responsive and informative in interpreting their inputs (Q3). However, P6 pointed out that the textual feedback could be overwhelming when it was too long, and suggested that incorporating visual cues could make it easier to understand, especially when their intended interactions were not supported.
All participants agreed that the system helped them save time when creating interactions (Q5), and most of them found the interaction creation process easy to use (Q4).
The two neutral scores for Q4 were given by P11, who suggested adding auto-complete functions, and P6, who required a preview feature before applying changes.

\textbf{Usability.}
Most participants agreed that they could quickly grasp how to use the system (Q6), indicating that the natural language interface and interaction framework were highly learnable. To further evaluate learning difficulty, we recorded the time each participant took to add their first interaction. Five participants (P2, P3, P6-P8) completed the task within 5 minutes, another five (P1, P4, P5, P9, P10) within 5-10 minutes, and only one participant (P11) took more than 10 minutes.
Participants also expressed a strong interest in continued use of the system (Q7) and would recommend it to others (Q8).

\textbf{Open-ended Questions.}
In response to Q9, all participants highlighted the natural language interface for adding interactions as the most useful feature of the system. Even the two visualization experts (P2 and P5) emphasized that while coding offered greater flexibility, the natural language interface was more intuitive and easier to use for creating most types of interactions.
Several participants also highlighted the specific strengths of the system. 
P9 commented, ``\textit{When my description is not clear enough, the system provides similar suggestions, which makes it easier for me to find the interaction I need.}''
P4 appreciated the responsiveness, stating, ``\textit{The feedback is instant, so I can immediately tell whether the change is what I intended.}''
P8 mentioned, ``\textit{Directly filtering out unwanted elements from the chart is incredibly convenient. Previously, I often had to rebuild the entire visualization to achieve the same result.}''
In response to Q10, participants also suggested areas for improvement. P1 and P5 mentioned that the system's processing time for adding interactions could be faster. P11 suggested that the system should provide more examples at the beginning when the user is not familiar with the system.
For future improvements (Q11), P3 and P9 expressed a desire for customized chart types like composite charts. P6 and P11 suggested adding voice command functionality to support hands-free interaction. P8 proposed the inclusion of a customizable color palette to enable more flexible styling of visualization elements.

%% file: chapter/8_discussion.tex
\section{Discussion and Future Work}
Our approach transforms static visualizations into interactive ones through natural language requirements.
This approach enabled interactivity in many previously static visualizations. However, several important considerations and limitations warrant discussion.

\textbf{Scope of Interaction Support.}
The current approach derives data exclusively from the existing visualization.
Consequently, the supported interaction modes are limited to manipulating information already present in the visualization through rearrangement.
However, many interactions that require external data, such as drill-down operations where users seek more detailed information about elements of interest, exceed the capabilities of our current approach.
When users request information beyond what can be inferred from the visual representation, the approach cannot respond adequately.
Future work necessitates the integration of external data binding mechanisms to support these types of interactions.


\revision{\textbf{Scalability and Adaptive Parsing.}
Our framework decouples the interaction logic synthesis from the visual rendering scale.
For visualizations with high-cardinality elements, such as area charts containing thousands of data points, the multi-agent requirement analyzer focuses on structural patterns rather than individual visual mark, ensuring that logical complexity does not increase with data density.
Currently, thousands of data points represent a typical upper bound for maintaining performance in web-based interactive visualizations, which remains within the context window limits of state-of-the-art MLLMs.
To further ensure robustness, our system employs an adaptive parsing strategy: if the element count exceeds a safe threshold, the system can pivot to extracting only essential metadata and text elements (e.g., axis labels, legends, and titles).
This allows the MLLM to infer the chart's semantic structure (axes and legends) and intent without being overwhelmed by excessive tokens.}

\revision{\textbf{Generalizability and Chart Types.}
Currently, \toolname{} primarily supports common statistical visualizations within the Cartesian coordinate system. 
Bar charts (simple, grouped, stacked), line charts, area charts (simple, overlapped, stacked), and scatter plots, are generally regarded as the most frequently used charts~\cite{battle2018beagle}, which are all well-supported by our MLLM-based parsing and interaction mapping.
However, our approach faces challenges with visualizations that require complex spatial layout computations or specialized data structures. For instance, force-directed graphs, geographic maps, and volumetric visualizations often involve non-linear spatial mappings or intricate topological relationships that are difficult to reconstruct purely from static SVG elements without underlying data models. Future work could involve enhancing the multi-agent reasoning capabilities to better infer these complex layouts or integrating lightweight geometry solvers to support a broader range of non-Cartesian visualizations.}

\revision{\textbf{Input Formats and Core Contributions.}
Our current implementation primarily operates on SVG-based visualizations, as our core contribution lies in the downstream interaction-authoring pipeline rather than the initial parsing of visual structures. SVG serves as a structured intermediate representation that enables the system to focus on the mapping from visual elements to interactive behaviors. Extending the system to support bitmap visualizations is an independent and decoupled process, which could utilize advanced MLLM-based techniques to extract visual hierarchies. For instance, recent work like SimVecVis~\cite{liu2025simvec} demonstrates how LLMs can effectively reconstruct visual elements and spatial structures from bitmap images. Since this parsing stage is decoupled from our core interaction-generation logic, future work could integrate such modules to make our framework format-agnostic.}

\section{Conclusion}
We present \toolname{} for adding interactive capabilities to static visualizations based on natural language specifications.
Our action-modification interaction design space provides a structured framework, abstracting interactions into user actions and corresponding visualization modifications.
A multi-agent requirement analyzer translates natural language instructions into actionable specifications, accommodating diverse user needs without technical expertise.
An implementation-agnostic representation translator converts static visualizations into flexible representations that can be manipulated independently of their original implementation.
Evaluation results show that our approach effectively adds interactive functionalities to common chart types while preserving high visual fidelity to the original visualizations.